\documentclass[iop]{emulateapj}
\usepackage[colorlinks,urlcolor=blue,citecolor=blue,linkcolor=blue]{hyperref}

\shorttitle{Shock acceleration: diffusion}
\shortauthors{Caprioli \& Spitkovsky}

\begin{document}

\title{Simulations of Ion Acceleration at Non-relativistic Shocks.\\
III. Particle Diffusion}

\author{D. Caprioli and A. Spitkovsky}
\affil{Department of Astrophysical Sciences, Princeton University, 
    4 Ivy Ln., Princeton NJ 08544}
\email{caprioli@astro.princeton.edu}

\begin{abstract}
We use large hybrid (kinetic protons--fluid electrons) simulations to investigate the transport of energetic particles in self-consistent electromagnetic configurations of collisionless shocks.
In previous papers of this series, we showed that ion acceleration may be very efficient (up to 10-20\% in energy), and outlined how the streaming of energetic particles amplifies the upstream magnetic field.
Here, we measure particle diffusion around shocks with different strengths, finding that the mean free path for pitch-angle scattering of energetic ions is comparable with their gyroradii calculated in the self-generated turbulence.
For moderately-strong shocks, magnetic field amplification proceeds in the quasi-linear regime, and particles diffuse according to the self-generated diffusion coefficient, i.e., the scattering rate depends only on the amount of energy in modes with wavelengths comparable with the particle gyroradius.
For very strong shocks, instead, the magnetic field is amplified up to non-linear levels, with most of the energy in modes with wavelengths comparable to the gyroradii of highest-energy ions, and energetic particles experience Bohm-like diffusion in the amplified field.
We also show how enhanced diffusion facilitates the return of energetic particles to the shock, thereby determining the maximum energy that can be achieved in a given time via diffusive shock acceleration.
The parametrization of the diffusion coefficient that we derive can be used to introduce self-consistent microphysics into large-scale models of cosmic ray acceleration in astrophysical sources, such as supernova remnants and clusters of galaxies.
\end{abstract}


\section{Introduction}

This paper is the third in a series of works aimed to investigate ion acceleration in non-relativistic collisionless shocks via large hybrid (kinetic ions--fluid electrons) simulations.
In previous papers we discussed how diffusive shock acceleration \citep[DSA, e.g.,][]{bell78a,blandford-ostriker78} at strong shocks can be very efficient in accelerating particles \citep[][hereafter Paper I]{DSA}, and how energetic ions induce magnetic field amplification via plasma instabilities \citep[][Paper II]{MFA}.

The connection between magnetic field amplification and particle acceleration is prominent in supernova remnants (SNRs), which are regarded as the sources of Galactic cosmic rays (CRs) up to the so-called \emph{knee} (a few $Z$ PeV, with $Z$ the nucleus charge). 
Our kinetic simulations support such a paradigm:
in Paper I, we found that shocks propagating almost along the large-scale magnetic field $\bf B_0$ (quasi-parallel shocks)  channel 10--20\% of their bulk flow energy into energetic ions. 
In Paper II, we showed that, when acceleration is efficient, the initial magnetic field is effectively amplified.
The total magnetic field is found to scale as $B_{tot}/B_0\approx \sqrt{M_A/2}$, where $M_A=v_{sh}/v_A$ is the Alfv\'enic Mach number, i.e., the ratio of the shock velocity $v_{sh}$ and the Alfv\'en speed $v_A=B/\sqrt{4\pi m n}$ (with $m$ and $n$ the proton mass and number density).
These results are consistent with multi-wavelength observations of young remnants, which suggest that magnetic fields at SNR blast waves are several tens to hundred times larger than in the interstellar medium \citep[see, e.g.,][]{P+06,Uchiyama+07,tycho,reynoso+13}.

DSA predicts that strong shocks should accelerate particles with a universal power-law  $\propto p^{-4}$ in momentum, and in Paper I we confirmed such scaling for the first time in kinetic simulations.
In spite of DSA spectral slope being independent of the details of particle scattering, magnetic field amplification regulates particles diffusion, and thereby the acceleration rate and the maximum energy that can be achieved.
The acceleration time up to energy $E$ is $T_{acc}\approx D(E)/v_{sh}^2$, where $D(E)$ is the spatial diffusion coefficient \citep[see, e.g.,][]{drury83}.
This characteristic time must be compared with the duration of the ejecta-dominated stage, since in the Sedov stage shock velocity and magnetic field amplification drop quite rapidly, and highest-energy particles are expected to escape the accelerator \citep[see, e.g.][]{escape}.

In order for SNRs to accelerate CRs up to the knee, diffusion near a shock has to be dramatically enhanced compared to diffusion in the interstellar medium.
From the B/C ratio in CRs, one infers an average Galactic diffusion coefficient $D_G(E)\approx 2\times 10^{28} (E/10 {\rm GeV})^{1/3}{\rm cm}^2{\rm s}^{-1}$ \citep[see, e.g.,][]{ba11a}.
However, if the diffusion coefficient at SNR shocks were not much smaller than $D_G(E)$, the maximum energy achievable in these sources would be less than $\sim 10$GeV.
This follows from comparing the acceleration time $T_{acc}\simeq D_G(E_{max})/v_{sh}^2$ with the Sedov time $T_{Sed}\sim 10^3$yr, when $v_{sh}(T_{Sed})\approx 4000$kms$^{-1}$.
Acceleration is significantly faster if the mean free path for pitch-angle scattering is as small as the particle gyroradius, $r_L$ (usually referred to as \emph{Bohm diffusion}), in which case the diffusion coefficient reads $D_B(E)\approx v r_L(E)\propto E/B$, where $v$ is the particle velocity. 
Still, for Bohm diffusion in the typical Galactic field  of a few $\mu$G, one obtains $E_{max}\approx 10^4-10^5$GeV, more than an order of magnitude below the CR knee \citep[e.g.,][]{lagage-cesarsky83b}.
A natural explanation is that the magnetic field amplification inferred in young remnants may be responsible for enhancing ion scattering beyond the Bohm limit calculated in the Galactic field: in this case, the actual diffusion coefficient would be Bohm {\em in the amplified magnetic field}, i.e., a factor of $\delta B/B_0\sim 10-100$ smaller.

In this paper we characterize the transport of energetic ions in kinetic simulations of non-relativistic collisionless shocks, in which the magnetic irregularities responsible for particle scattering are generated self-consistently by the different flavors of streaming instability.
The paper is structured as follows. 
In Section \ref{sec:hybrid} we summarize the properties of the self-generated magnetic turbulence, as inferred from the large simulations of Paper II.
In Section \ref{sec:diff} we point out a novel technique inspired by analytical approaches to DSA, which returns the mean diffusion coefficient relevant for energetic particles in the shock precursor;
our findings are supported by the analysis of individual ion tracks, which provides the spatial dependence of the diffusion coefficient in different shock regions (Section \ref{sec:track}).
Throughout the paper, we interpret our results in terms of Bohm and self-generated diffusion coefficient.
We discuss how diffusion regulates the time evolution of the maximum energy achieved by accelerated ions in Section \ref{sec:Emax}, and conclude in Section \ref{sec:concl}.

\section{Hybrid simulations}\label{sec:hybrid}
\begin{table}
  \caption{Parameters of the relevant hybrid runs (see also Paper II)}\label{tab:box}
  \begin{center}
    \begin{tabular}{cccccc} \hline \hline              
  Run & M & $x$ $[c/\omega_p]$  & $y$ $[c/\omega_p]$ & $t_{max}[\omega_c^{-1}]$ & $\Delta t [\omega_c^{-1}]$\\ 
  \hline 
  A	  & 20 & $5\times 10^4$	& $1000$ & $1000$ & $5\times 10^{-4}$ \\
  B	  & 20 & $ 10^5$	& $100$ & $2500$ & $5\times 10^{-4}$  \\
  D 	  & 80 & $ 4\times 10^5$ & $200$ & $500$ & $ 2.5\times 10^{-4}$  \\
  F 	  & 60 & $ 2\times 10^5$ & $20$ & $1600$ & $ 2.5\times 10^{-4}$  \\
  \hline
    \end{tabular}
  \end{center}
\end{table}
In Paper I and II we have discussed simulations of non-relativistic, collisionless shocks performed with the \emph{dHybrid} code \citep{gargate+07}.
The main strength of hybrid simulations, which treat ions as kinetic particles, and electrons as a neutralizing fluid \citep[see, e.g.,][]{lipatov02}, is to allow for larger/longer simulations (in physical units) with respect to full particle-in-cell methods.
We measure lengths in units of ion skin depth $c/\omega_p$, where $\omega_p=\sqrt{4\pi n e^2/m}$ is the ion plasma frequency, and time in units of inverse cyclotron frequency $\omega_c^{-1}=mc/eB_0$, with $c$ the speed of light and $e$ the ion charge.
Velocities are normalized to the Alfv\'en speed $v_A=B_0/\sqrt{4\pi m n}$, and energies to $E_{sh}=\frac m2 v_{sh}^2$, where ${\bf v}_{sh}= -v_{sh}{\bf x}$ is the velocity of the upstream fluid in the simulation frame.
The shock is characterized by its Alfv\'enic Mach number $M_A=v_{sh}/v_A$, and throughout the paper we assume the sonic Mach number to be roughly equal to $M_A$ (thermal to magnetic pressure ratio $\beta=2$), indicating both simply with $M$.

Among the runs in Paper II, we focus on the ones summarized in table \ref{tab:box}.
They correspond to strong shocks ($M\gg 1$), and account for different features: large transverse size (Run A), long term evolution (Run B), and very large $M$ (Run D).
Our unprecedentedly-large 2D and 3D simulations allow us to assess that ions are accelerated via DSA at quasi-parallel shocks (${\bf v}_{sh}$ almost parallel to ${\bf B}_0$), and that the initial magnetic field is amplified in the precursor up to
\begin{equation}\label{eq:db}
\left\langle\frac{B_{tot}}{B_0}\right\rangle^2\approx \frac{M}{2},
\end{equation}
if the acceleration efficiency is $\zeta_{cr}\approx 15\%$ (Equation~2 in Paper II).
The spectrum of the excited turbulence depends on the shock strength, as illustrated in Section 4 of Paper II.
For $M\lesssim 30$, the turbulence spectrum is consistent with the quasi-linear prediction of resonant streaming instability \citep[e.g.,][]{skilling75a, bell78a}, while for stronger shocks the non-resonant hybrid \citep[NRH,][]{bell04,bell05} instability is the fastest to grow.
In the high-Mach number regime, the turbulence is excited far upstream by the streaming of escaping ions (with energy close to the maximum energy $E_{max}$), and the most unstable mode has wavenumber $k_{max}\gg 1/r_L(E_{max})$.
In the nonlinear stage of the instability (when $b\equiv\delta B/B_0\gg 1$), one has $k_{max}(b)\propto b^{-2}$ \citep[see also][]{rs09}, and the condition $kr_L(E_{max})\approx 1$ may be met at some point in the precursor. This resonance disrupts the coherence of the ion current, and leads to pitch-angle diffusion of energetic ions (see Section 5 in Paper II for more details).
We define the CR precursor as the upstream region where ions with energy up to $\sim E_{max}$ diffuse, and the far upstream as the region where the most energetic CRs escape freely, triggering the NRH instability.
Note that, since typically in the precursor  $b\gg1$, the helicity of the self-generated waves is unimportant, and it makes little sense to distinguish between resonant and NRH instability.
We argue that this is the reason why Equation~\ref{eq:db}, which matches the prediction of resonant streaming instability \citep[e.g.,][]{ab06}, provides a good description of the maximum amplification factor achieved in the upstream even for very strong shocks (figure 5 in Paper II).

\section{Diffusion coefficient}\label{sec:diff}
Quantitative approaches to DSA are based on a description of the CR transport, which typically requires an a priori description of how particles diffuse while being advected with the fluid \citep[see][for a comparison of numerical, Monte Carlo and analytical approaches to non-linear DSA]{comparison}.
The simplest description introduces a diffusion coefficient accounting for ion diffusion parallel to the local magnetic field \citep[see, e.g.,][and references therein for generalizations to anisotropic diffusion]{jokipii87,Bill+03}.
Even Monte Carlo approaches \citep[e.g.][]{jones-ellison91}, which do not adopt an explicit diffusion coefficient, need to prescribe the CR mean free path for pitch-angle scattering.
In this Section, we calculate CR diffusion in our kinetic simulations, in order to study the feedback of self-generated turbulence on energetic particles.

\subsection{Bohm and self-generated diffusion}
For non-relativistic shocks, the most popular choice is to assume that particles diffuse via small-angle deflections, with a mean free path of the order of the particles' gyroradius $r_L$ (Bohm diffusion).
The corresponding diffusion coefficient reads\footnote{Note the factor of 2 in the denominator, which is peculiar of the reduced space-dimensionality of the simulation: it is equal to 1 in 1D, to 2 in 2D, and to the canonical value of 3 in 3D.}: 
\begin{equation}\label{eq:DB}
D_B(E)\equiv\frac{v}{2}r_L(v,B)=\frac{v}{2}\frac{pc}{eB}=\frac{E}{m\omega_c}\,.
\end{equation}
In this limit, the spectral energy distribution of the magnetic irregularities is neglected, while in reality one must consider that ions with momentum $p$ preferentially scatter against modes with wavenumber $\bar{k}_p=m\omega_c/p$. 
Strictly speaking, this resonance condition should involve only the component of ${\bf p}$ along ${\bf k}\parallel \hat{\bf x}$, and also a multiplicative factor of $\mathcal{O}(1)$, which can be ignored for our purposes \citep[see, e.g.,][for details]{skilling75c}. 

As in Paper II, we introduce the normalized magnetic energy density per unit logarithmic bandwidth of waves with wavenumber $k$, $\mathcal{F}(k)$, defined as: 
\begin{equation}\label{eq:F}
\frac{B_{\perp}^2}{8\pi}=\frac{B_{0}^2}{8\pi}\int_{k_{min}}^{k_{max}}\frac{dk}{k}\mathcal{F}(k),
\end{equation}
where $B_{\perp}$ is the transverse component of the magnetic field, $\mathcal{F}(k)/k=|\tilde{B}_y(k)|^2+|\tilde{B}_z(k)|^2$, and $\tilde{B}_{i}(k)$ is the Fourier transform of $B_{i}(x)$.
In the quasi-linear limit, the diffusion coefficient in the presence of Alfv\'enic modes with spectrum $\mathcal{F}(k)$ reads \citep[see, e.g.,][]{bell78a}:
\begin{equation}
D(p)=\frac{4}{\pi}\frac{v(p)}{3}\frac{r_L(p)}{\mathcal{F}(\bar{k}_p)}.
\end{equation}
By using Equation~\ref{eq:DB}, and considering that the magnetic turbulence is produced by energetic ions themselves, we can cast the \emph{self-generated} diffusion coefficient as:
\begin{equation}\label{eq:Dsg}
D_{sg}(E)=\frac{8}{3\pi}\frac{D_B(E)}{\mathcal{F}(\bar{k})},
\end{equation}
where $\bar{k}$ is the resonant wavenumber. 
Equation~\ref{eq:Dsg} emphasizes how $D_B$ corresponds to pitch-angle diffusion in Alfv\'enic magnetic turbulence with $\mathcal{F}(k)\approx 1$  at all wavelengths, and also to $B_\perp/B_0\approx 1$, according to Equation~\ref{eq:F}. 

In Paper II we showed that, for $M=20$, the shock generates a $p^{-4}$ distribution of non-relativistic particles that excites a wave spectrum $\mathcal F(k)\propto k^{-1}\propto p$ (figure 6 of Paper II); 
therefore, the self-generated diffusion coefficient scales as $D_{sg}(p)\propto E/p\propto p$ (while $D_B\propto E$, instead).
The peculiar $\mathcal F\propto k^{-1}$ scaling depends on the fact that our CRs are non-relativistic: for a $p^{-4}$ distribution of relativistic CRs, in the presence of resonant streaming instability, one would have a flat $\mathcal F(k)\approx \mathcal F_0$ distribution, and the suppression of the diffusion coefficient would be independent of $p$ and proportional to $\mathcal{F}_0\propto \delta B/B_0$.   
Nevertheless, the different scaling of $\mathcal F(k)$ in the relativistic and in the non-relativistic regime compensates the corresponding scaling of $D_B(E)$, in such a way that $D_{sg}(p)\propto p$ is realized \emph{at any momentum}.

\subsection{Extracting the diffusion coefficient from simulations}\label{sec:anal} 
Let us consider the stationary, one-dimensional diffusion-convection equation for the isotropic (in momentum space) distribution function of accelerated particles, $f(x,p)$, \citep[e.g.,][]{skilling75a}:
\begin{equation}\label{eq:convdiff}
u\frac{\partial f}{\partial x} =  \frac{\partial }{\partial x}\left[ D(x,p)\frac{\partial f}{\partial x}\right]+\frac{p}{3}\frac{{\rm d} u}{{\rm d} x}\frac{\partial f}{\partial p}\,.
\end{equation}
With the boundary condition $f(p)\to 0$ at upstream infinity, an excellent approximate solution of the equation above reads \citep[see, e.g.,][]{boundary}: 
\begin{equation}\label{eq:fxp}
f(x,p)=f_{sh}(p)\exp{\int_0^x{\rm d}x' \frac{u(x')}{D(x',p)}},
\end{equation}
where 
\begin{equation}
f_{sh}(p)\propto \left(\frac{p}{p_{inj}}\right)^{-q};\quad q=\frac{3r}{r-1}
\end{equation}
is the CR distribution function at the shock, and $r$ is the shock compression ratio. 
All the equations are written in the shock reference frame\footnote{Our simulations are in the downstream reference frame, instead.}: 
the shock is at $x=0$, the upstream is for $x>0$, and $u(x)<0$ is the fluid velocity (we dropped the notation $\tilde{u}$ we adopted in Paper II).

\begin{figure}\centering
\includegraphics[trim=0px 70px 10px 0px, clip=true, width=.485\textwidth]{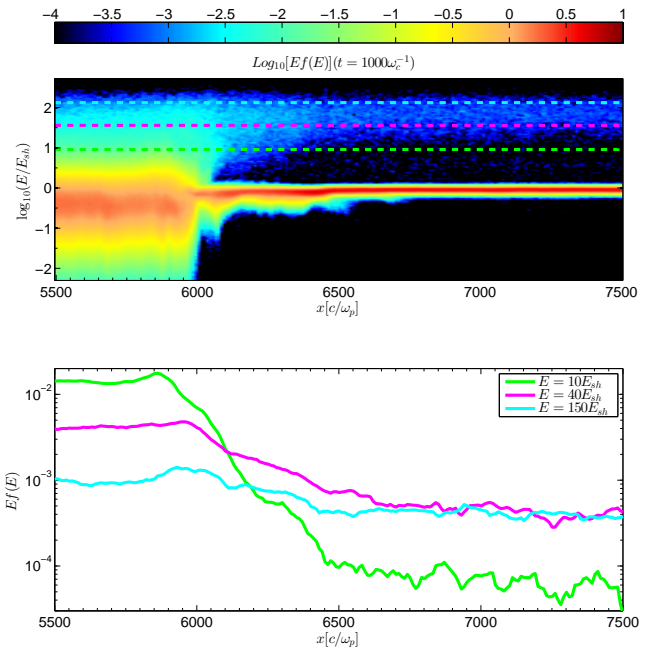}
\caption{\label{fig:fxp}
\emph{Top panel}: ion spectrum (color code) at different positions around the shock with $M=20$ (Run A), at $t=1000\omega_c^{-1}$.
The upstream cold beam is thermalized at the shock front ($x_{sh}\approx 6000c/\omega_p$), while ions with $E\gtrsim E_{sh}$ diffuse upstream of the shock, proportional to their energy.
\emph{Bottom panel}: differential density profile of CRs corresponding to the dashed lines in the top panel.
Low-energy CRs are confined close to the shock, while more energetic ions diffuse much further.
High-energy ions also show a shallow jump across the shock, their gyroradii being much larger than the shock thickness. 
Note that the CR distribution is increasingly dominated by higher-energy particles further into the upstream.  
\emph{A color figure is available in the online journal.} }
\end{figure}

In Paper I we have shown that the CR spectrum at the shock is consistent with the DSA prediction;
now we want to check that also the expected spatial dependence of the upstream CR  distribution is recovered.
Figure \ref{fig:fxp} shows the ion spectrum as a function of position (top panel), for a parallel shock with $M=20$ (Run A).
The bottom panel in the same figure illustrates the distribution $Ef(x,E)$ for accelerated particles with energy $E=10,40,150 E_{sh}$, corresponding to the dashed lines in the top panel.
Three things are worth noting.
First, the larger the energy, the larger the extent of the distribution upstream of the shock. 
At any position in the precursor, the CR spectrum has a low-energy cut-off, which moves to higher energies for larger $x$, in qualitative agreement with Equation~\ref{eq:fxp}. 
Second, the CR distribution at higher energies has a smaller jump across the shock (see, e.g., the curve for $150E_{sh}$ in Figure \ref{fig:fxp}), much smaller than the compression factor $r\approx 4$. 
This behavior is peculiar of ions with gyroradii larger than the shock thickness, and induces a non-linear modification of shock jump conditions, as discussed in Paper I.
Third, moving from the shock towards the upstream, the CR distribution is first exponentially suppressed, and then flattens at a level $\ll f_{sh}(p)$ \citep[as also observed by][]{giacalone04}.
As discussed in Section 5 of Paper II, the diffusion approximation breaks far upstream, and for $E\gtrsim E_{max}$, where the fraction of escaping ions is close to unity and the CR spectrum is cut off.
In these cases, care should be taken when interpreting results obtained by using Equation~\ref{eq:convdiff}.

The diffusion coefficient describing the self-generated ion scattering in the precursor can be worked out by considering the upstream distribution of non-thermal particles.
Let us assume that $D(x,p)$ and $u$ are constant in $x$, at least where $f(x,p)\simeq f_{sh}(p)$, so that in the upstream $f(x,p)\simeq f_{sh}(p)\exp [ux/D(p)]$. 
By integrating Equation~\ref{eq:fxp} from 0 to a given $x_0$, arbitrarily chosen such that $f(x_0,p)\ll f_{sh}(p)$, we get:
\begin{equation}
F_{up}(p)\equiv\int_0^{x_0} f(x,p) dx\simeq f_{sh}(p) \int_0^{x_0} \exp\left[\frac{ux}{D(p)}\right]dx\,,
\end{equation}
from which
\begin{equation}\label{eq:cd}
D(p)\simeq \frac{uF_{up}(p)}{f_{sh}(p)}\,.
\end{equation}
$F_{up}(p)$ is a global quantity that can be calculated easily in simulations, and we checked that results do not strongly depend on the particular choice of $x_0$, and on timescales shorter than the dynamical ones.
We comment more on these points in Section \ref{sec:track}.

\begin{figure}\centering
\includegraphics[trim=55px 0px 30px 30px, clip=true, width=.5\textwidth]{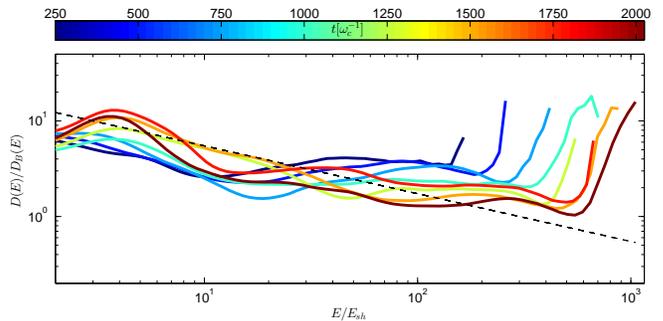}
\caption{\label{fig:CoeffDiff20}
Time evolution of the diffusion coefficient in a parallel shock with $M=20$ (Run B), calculated via Equation~\ref{eq:cd} with $x_0=10^4c/\omega_p$, and plotted as divided by the Bohm diffusion coefficient (Equation~\ref{eq:DB}).
At late times, normalization and energy scaling match well the self-generated diffusion coefficient (dashed line), which corresponds to Equation~\ref{eq:Dsg} with $\mathcal F=1$ at $k=1/r_L(E_{max}$) and $E_{max}\approx 300E_{sh})$.
\emph{A color figure is available in the online journal.}}
\end{figure}

\begin{figure}\centering
\includegraphics[trim=55px 0px 30px 30px, clip=true, width=.5\textwidth]{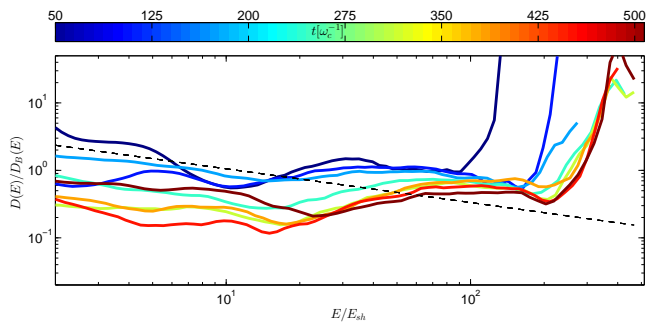}
\caption{\label{fig:CoeffDiff80} As in Figure \ref{fig:CoeffDiff20}, for a parallel shock with $M=80$ (Run D) and $x_0=2\times 10^4c/\omega_p$.
The inferred diffusion coefficient, at late times, is smaller than Bohm by a factor $\lesssim 5$, consistent with the level of magnetic field amplification in the precursor ($\delta B/B_0\approx 3-5$, see figure 7 in Paper II).
The dashed line corresponds to Equation~\ref{eq:Dsg} with $\mathcal F=3$ at $k=1/r_L(E_{max}$) and $E_{max}\approx 100E_{sh})$.
\emph{A color figure is available in the online journal.}}
\end{figure}

The diffusion coefficient estimated with the procedure above is plotted in Figure \ref{fig:CoeffDiff20} for Run B ($M=20$), as a function of time, and normalized to $D_B(E)$.
The inferred diffusion coefficient is a factor of few larger than Bohm in the background field, but at later times, when self-generated fields have had sufficient time to grow, it becomes comparable to $D_B$ close to $E_{max}$.
The energy dependence of $D$ agrees well with the self-generated diffusion coefficient, i.e., $D_{sg}(p)\propto p$ (Equation \ref{eq:Dsg} with $\mathcal F(k)\propto k^{-1}$, as generated via streaming instability by a $f(p)\propto p^{-4}$ distribution of non-relativistic particles).
The dashed line corresponds to Equation~\ref{eq:Dsg}, normalized by posing $\mathcal F\approx 1$ at $k=1/r_L(E_{max}$), where $E_{max}\approx 300E_{sh})$;
such a normalization is consistent with the wave spectrum $\mathcal F(k)$ in the precursor (see figure 6 in Paper II), and is comparable with $D_B$ close to $E_{max}$ because $\mathcal F\approx 1$ at resonance with $E_{max}$, i.e., highest-energy ions feel $\delta B/B\approx 1$ on their gyration scales.
Finally, above $E_{max}(t)$, the diffusion coefficient increases quite rapidly because the lack of long-wavelength modes makes scattering very ineffective;
in this regime, however, the use of the diffusion--convection equation (Equation~\ref{eq:convdiff}) becomes questionable.

Equation~\ref{eq:Dsg} suggests that, when magnetic field amplification is effective ($\delta B/B_0\gtrsim 1$ and, in turn, $\mathcal{F}(k)\gtrsim 1$), the self-generated diffusion coefficient should be smaller than $D_B(E)$.
By repeating the exercise above for a very strong shock with $M=80$ (Run D), which shows high levels of magnetic field amplification,  we confirm this to be the case.
Figure \ref{fig:CoeffDiff80} illustrates the diffusion coefficient calculated by using Equation~\ref{eq:cd}: 
the measured diffusion coefficient is smaller than Bohm in the background field $B_0$ at nearly all energies, consistent with $\mathcal F(k)$ being larger than 1 at wavenumbers resonant with accelerated ions (see bottom panel of figure 7 in Paper II).

It is important to notice that the energy scaling of the diffusion coefficient does not follow the quasi-linear prediction for self-generated diffusion ($D_{sg}\propto p^{-1}$, dashed line in Figure \ref{fig:CoeffDiff80}), being apparently more consistent with Bohm diffusion.
The reason for this difference with respect to the $M=20$ case is that the spectrum of the magnetic turbulence for $M=80$ does not follow the $\mathcal F\propto k^{-1}$ trend, because of the relevance of the NRH instability.
As it follows from figure 7 in Paper II, most of the magnetic energy in the precursor is at scales comparable with the gyroradii of the most energetic ions, and such large-scale turbulence is found to effectively scatter particles of any energy.
Since for $M=80$ we find $\mathcal F\gg 1$, this result suggests how to extend the quasi-linear theory of self-generated diffusion (Equation~\ref{eq:Dsg}) into the regime of nonlinear field amplification.
The suppression of the diffusion coefficient can be quantitatively estimated as of the order of $\delta B/B_0$, which corresponds to \emph{Bohm diffusion in the total (amplified) field, at all the energies}.

\subsection{Comments and caveats}
First of all, we notice that when amplification is strongly nonlinear, the magnetic field becomes very tangled ($B_\perp/B_\parallel\sim 1$); 
in this regime the distinction between parallel and perpendicular diffusion \cite[e.g.,][]{jokipii87} is lessened, and isotropic spatial diffusion should provide a reliable approximation.  

Another result inferred from our simulations is that diffusion is enhanced even close to $E_{max}$.
This statement is independent of $M$, and relies on most of the wave energy being at wavelengths resonant with high-energy particles (figure 6, 7 in Paper II).
As discussed above, this effect is prominent for non-relativistic CRs, but it is expected even for non-too-steep relativistic distributions, the correction being just logarithmic for $f(p)\propto p^{-4}$.
In general, Bohm diffusion in $B_{tot}\sim b B_0$, with the large amplification factors $b\sim 10-30$ expected at very fast shocks (Paper II), should be fast enough to allow young SNRs to accelerate CRs up to the knee.
Nevertheless, more investigation of high-$M$ shocks are needed to definitively assess the ability of SNRs to act as PeVatrons.
We point out two main effects potentially contributing corrections to our findings:
i) the presence of filamentation \citep{filam}, which might lead to inhomogeneous diffusion; 
and ii) the limited duration (in physical time) of our simulations for large $M$.

Hints of Bohm-like diffusion in CR precursors have already been outlined by \cite{rb13}, who used a MHD-kinetic code that exploits a spherical harmonic expansion of the Vlasov--Fokker--Planck equation to calculate a self-consistent magnetic field configuration starting from an initial CR current of mono-energetic ions.
These authors inferred diffusion faster than Bohm in $B_0$, but slower than Bohm in $B_{tot}$, possibly because simulations were not converged (see their section 4.3).
Our hybrid simulations, instead, test CR diffusion for an ion power-law distribution that forms and grows spontaneously because of DSA, without any need to prescribe particle injection or escape; therefore, we can self-consistently study the connection between CR spectrum, wave spectrum, and momentum dependence of the diffusion coefficient.
We also distinguish between different regimes of field amplification, and confirm that for strong shocks Bohm diffusion in $B_{tot}$ provides a reasonable description for the transport of particles of any energy.
Finally, we attest to the relative relevance of NRH and resonant instability in amplifying the magnetic field in the far upstream and in the precursor, providing the theoretical framework for calculating both self-generated turbulence and diffusion for a given CR distribution, in different shock regions.

\section{Particle Tracking}\label{sec:track}
\begin{figure}\centering
\includegraphics[trim=0px 17px 0px 280px, clip=true, width=.5\textwidth]{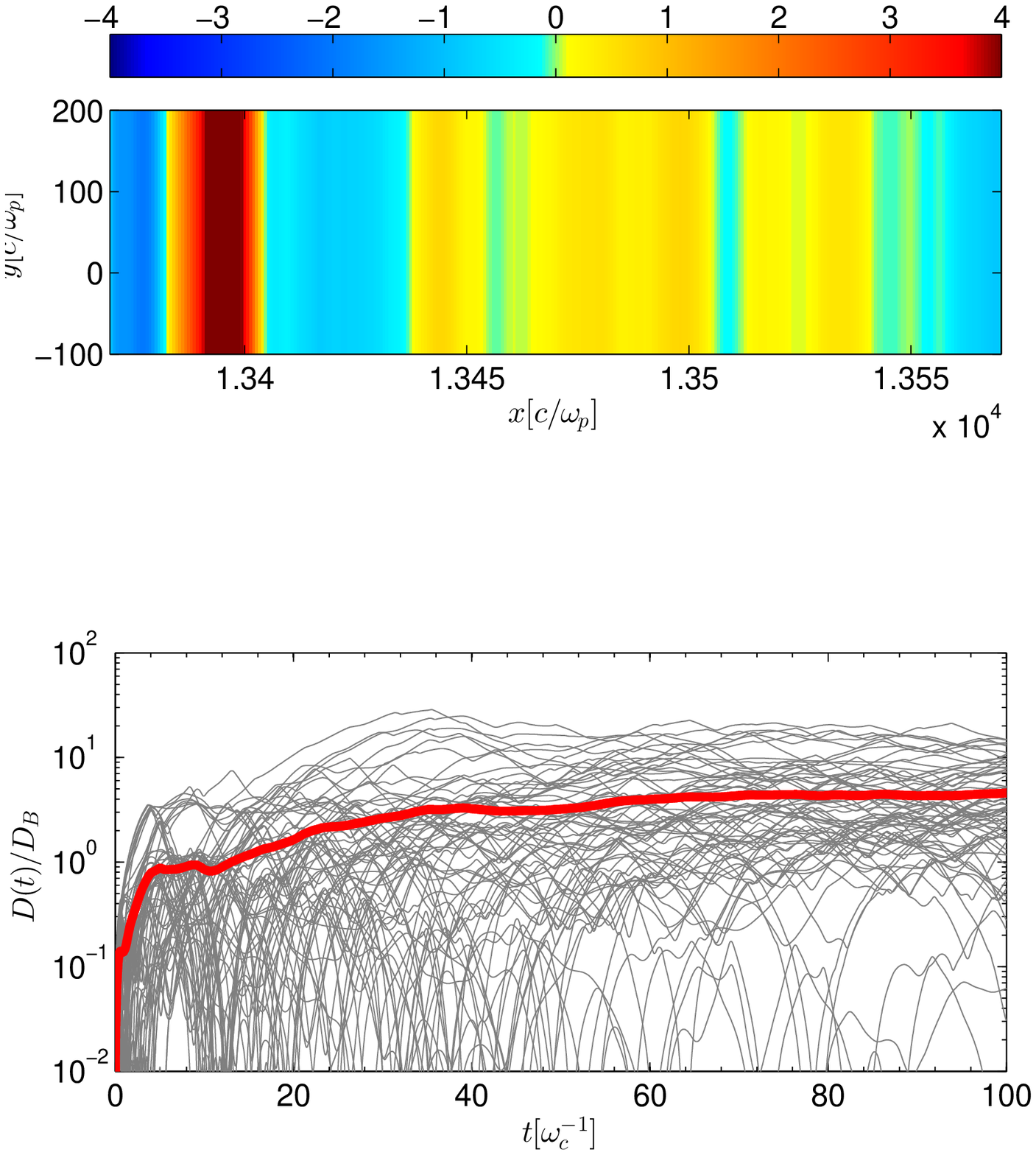}
\includegraphics[trim=0px 17px 0px 280px, clip=true, width=.5\textwidth]{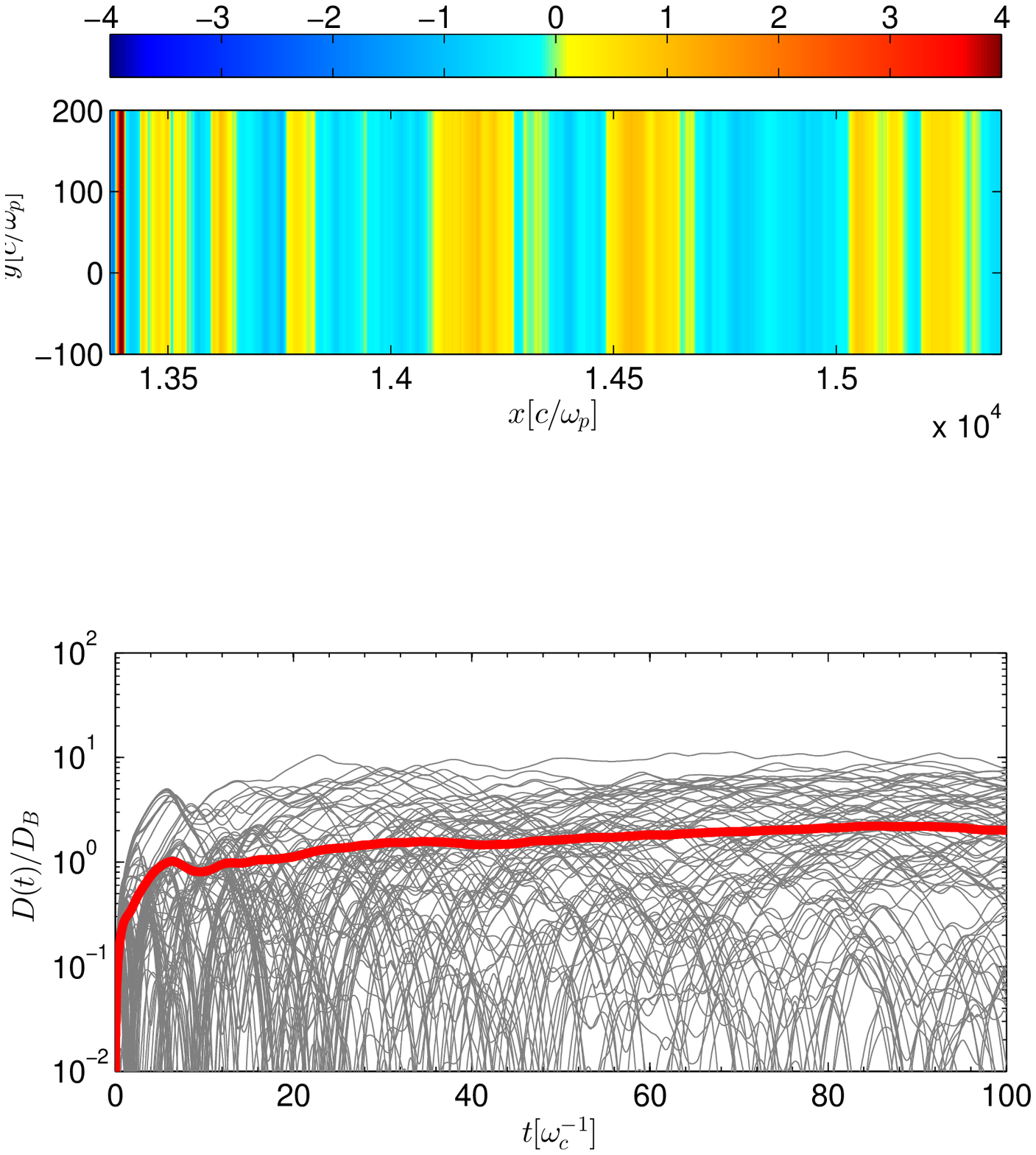}
\includegraphics[trim=0px 17px 0px 280px, clip=true, width=.5\textwidth]{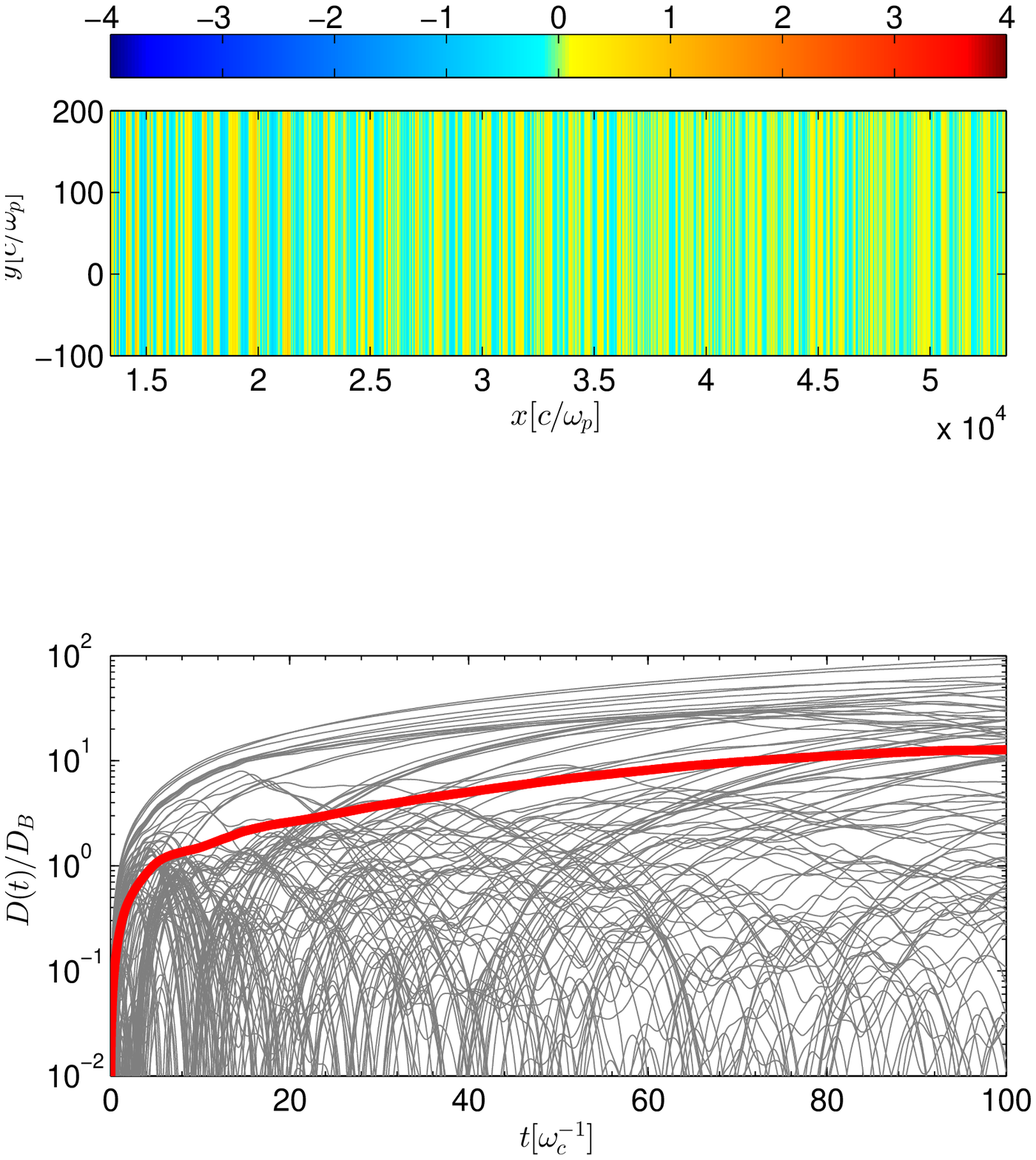}
\caption{\label{fig:track}
Running diffusion coefficient for $M=20$ shock (run B) at $t=2400\omega_c^{-1}$, in a region of width $\lambda(E)$ upstream of the shock. 
Gray curves show $D(t)$ for ensambles of 100 particles, with energy $E=10,100,1000E_{sh}$ (panels from top to bottom), and random initial velocity direction.
The red thick line in each panel shows the averaged diffusion coefficient (see Equation~\ref{eq:runD})
\emph{A color figure is available in the online journal.}}
\end{figure}

A natural question is whether the diffusion coefficient obtained with Equation~\ref{eq:cd} is recovered also when analyzing the trajectories of individual particles.
In order to reconstruct the  local diffusion coefficient, we select a region of the simulation box, take a snapshot of its electromagnetic configuration, impose periodic boundary conditions, and propagate many particles in it for long time using a Boris pusher \cite{boris70}.
Such an ``ergodic'' approach is valid if the shock structure does not vary dramatically on dynamical time-scales, and if fields are almost uniform in the box.
For any energy $E$, we use a box of width $\lambda(E)$, defined as twice the Bohm diffusion length in $B_0$, i.e.:
\begin{equation}\label{lambda}
\lambda(E)\simeq 2\frac{D_B(E)}{v_{sh}}=M\frac{E}{E_{sh}}\frac{c}{\omega_p}.
\end{equation}

The spatial diffusion coefficient along ${\bf B}_0$ can be calculated by taking the asymptotic time limit of the running diffusion coefficient $D(E,t)$, defined as:
\begin{equation}\label{eq:runD}
D(E)\equiv\lim_{t\to\infty} D(E,t) =\lim_{t\to\infty}\sum_{n=1}^N\frac{|x_n(t)-x_{n}(0)|^2}{2tN}.
\end{equation}
In order to remove statistical fluctuations, we average over a large number of particles, $N$, with fixed energy and random velocity direction.
Figure \ref{fig:track} shows the running diffusion coefficient in the precursor of run D (red thick line), averaged over $N=100$ random particles (thin gray lines) of different energies ($E=10,100,1000E_{sh}$, top to bottom panels, respectively).
For all the energies below $E_{max}\approx 300E_{sh}$, the running diffusion coefficient tends to an asymptotic value.
This confirms that particles are indeed diffusing, i.e., their mean displacement from the initial position $x(0)$ increases in time as $\langle \Delta x\rangle\propto \sqrt{2Dt}$.
Conversely, tracks of $E=1000E_{sh}$ particles (bottom panel of Figure \ref{fig:track}) show a mixed behavior: beside diffusing ions, we have free-streaming ions, for which $D(t)\propto t$ (see Equation~\ref{eq:runD});  
when averaged over all the particles, this bimodal distribution returns a large  $D\sim 10D_B$, but this result must be interpreted with a grain of salt.
Finally, we have some tracks that curve down for a while, which correspond to  ions momentarily trapped in the generated turbulence.

\begin{figure}\centering
\includegraphics[trim=0px 30px 0px 260px, clip=true, width=.485\textwidth]{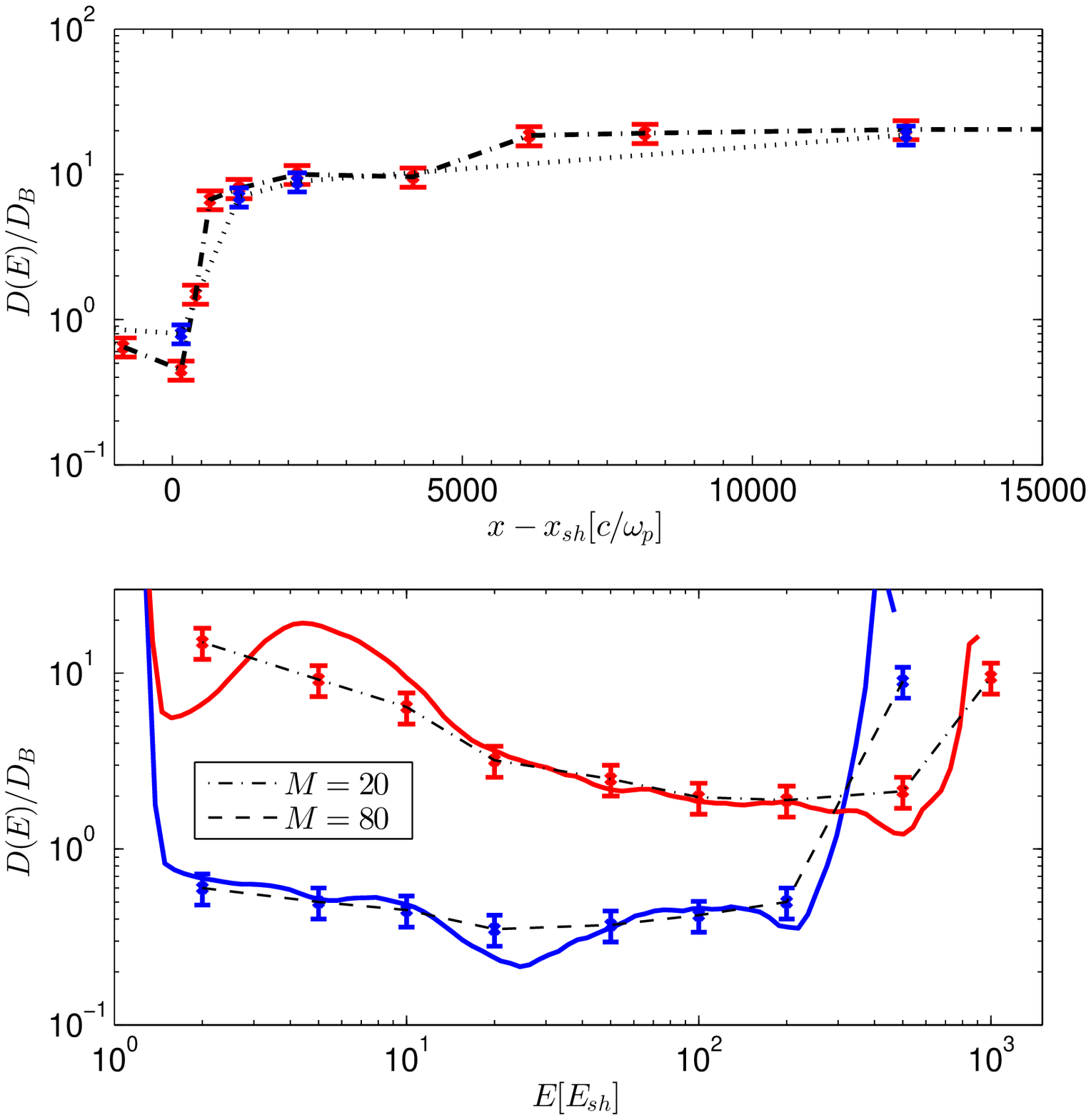}
\caption{\label{fig:cd2080}
Diffusion coefficient immediately in front of the shock for $M=20,80$, inferred by tracking individual particles (points with fiducial error bars of 20\%), and by using the procedure outlined in Section \ref{sec:anal} (solid red and blue lines, corresponding to the last time in Figures \ref{fig:CoeffDiff20} and \ref{fig:CoeffDiff80}).
Particles are propagated in periodic boxes with fields extracted from regions of thickness $\lambda(E)$ ahead of the shock.
\emph{A color figure is available in the online journal.}}
\end{figure}

In Figure \ref{fig:cd2080} we compare the upstream diffusion coefficient measured by averaging the CR distribution function over the upstream (as in Section \ref{sec:anal}, solid lines), and by tracking individual particles with different energies (symbols) in a region of thickness $\lambda(E)$ ahead of the shock in Runs B and D, at $t=2400\omega_c^{-1}$ and $t=500\omega_c^{-1}$, respectively.
The agreement between the two methods is very good, even at energies above $E_{max}$, in spite of Equation~\ref{eq:convdiff} becoming progressively less accurate. 

\begin{figure}\centering
\includegraphics[trim=0px 280px 0px 15px, clip=true, width=.485\textwidth]{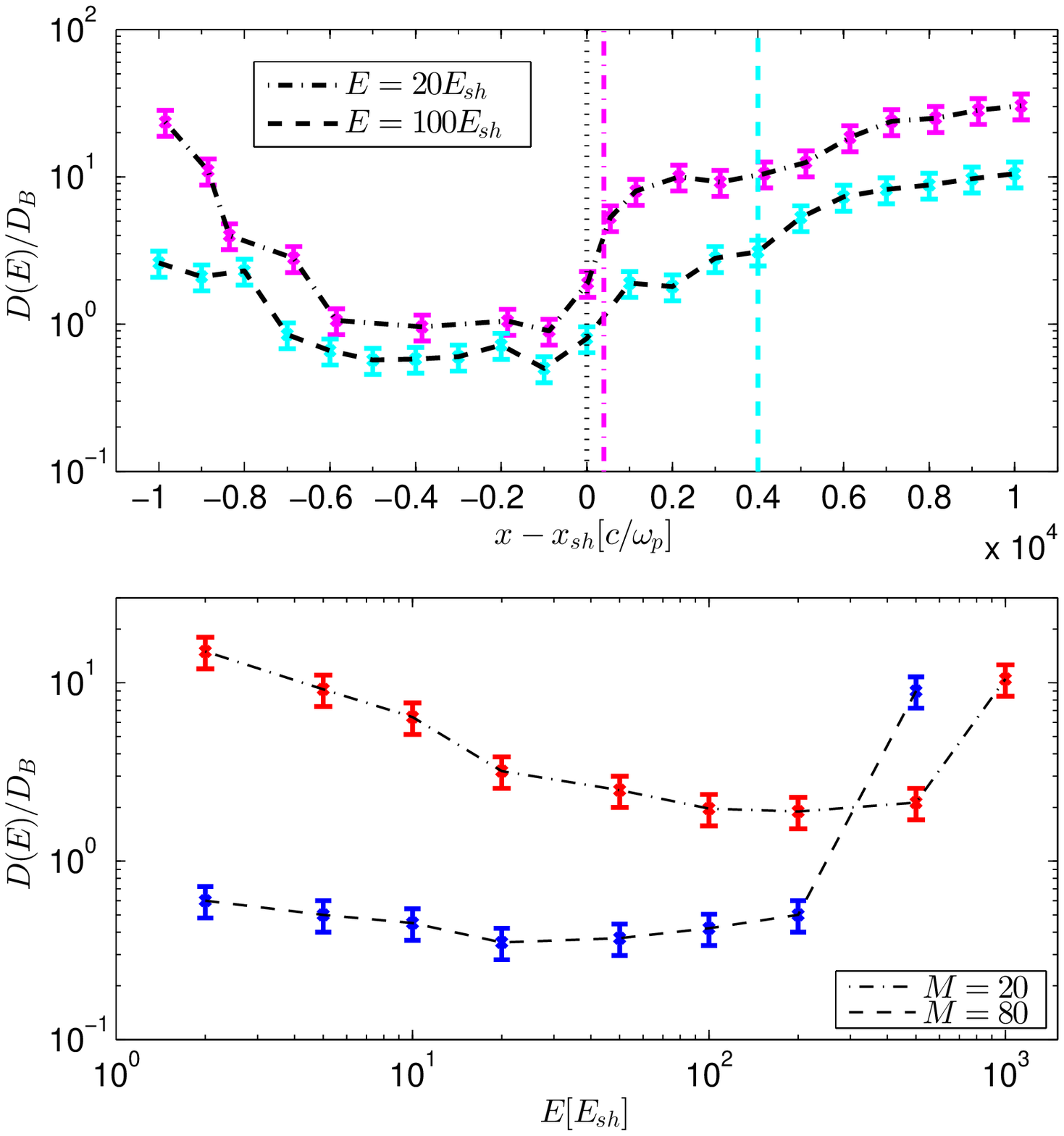}
\caption{\label{fig:cdx}
Spatial dependence of the diffusion coefficient for a Mach 20 shock (Run B), at $t=2400\omega_c^{-1}$, for ions with energy $E=20,100E_{sh}$ (magenta and cyan lines, respectively).
Points correspond to $D(E,x)$ calculated by tracking CRs in periodic boxes centered at $x$, and of width $2\lambda(E)$, indicated as the distance between the vertical colored lines and the dotted line; bars indicate a fiducial 20\% error. 
\emph{A color figure is available in the online journal.}}
\end{figure}

The biggest limitation of the analytical method of Section \ref{sec:anal} is that it only applies to the shock precursor;  
particle tracking is the only viable choice to study ion transport far upstream and in the downstream.
In Figure \ref{fig:cdx} we show the diffusion coefficient measured by propagating particles with energy $E=20,100E_{sh}$ in different regions of the shock in Run B. 
The spatial profile of the diffusion coefficient is rather similar at different energies, and shows a minimum behind the shock, where $D(E)$ is about $r\approx 4$ times smaller than immediately upstream, consistently with field compression at the shock.
Finally, the diffusion coefficient increases when moving toward upstream and downstream infinity, consistently with the profile of the self-generated turbulence.

\section{Maximum ion energy}\label{sec:Emax}

The diffusion rate is of primary importance for determining the maximum energy achievable via DSA.
Let us consider the evolution of $E_{max}(t)$, determined by fitting the post-shock non-thermal ion spectrum (figure 2 in Paper II) with a power-law $\propto E^{-1.5}$, plus an exponential cut-off at $E_{max}(t)$.
In DSA, the instantaneous maximum ion energy is regulated by the finite acceleration time, the other potential limiting factors  being the size of the system, which must encompass particle trajectories, or energy losses (usually relevant for leptons only).

The acceleration rate depends on the time it takes a particle to diffuse back and forth across the shock, and is calculated as \citep[see, e.g.,][]{drury83}: 
\begin{equation}\label{eq:Tacc}
T_{acc}(E)=\frac{3}{u_1-u_2}\left[\frac{D_1(E)}{u_1}+\frac{D_2(E)}{u_2} \right],
\end{equation} 
where the subscripts 1 and 2 refer to upstream and downstream, respectively.
For simplicity, we assume $u$ (the fluid velocity in the shock reference frame) and $D$ to be piecewise constant upstream and downstream; Equation~\ref{eq:Tacc} can be generalized to the case of efficient CR acceleration, in which these quantities depend on $x$ \citep{BAC07}. 
We then pose $D_1\simeq r D_2\equiv D$ (as inferred from Figure \ref{fig:cdx}), and rewrite Equation \ref{eq:Tacc} as:
\begin{equation}\
T_{acc}(E)\simeq\frac{6r^3}{(r^2-1)(r+1)}\frac{D(E)}{v_{sh}^2}\simeq\frac{6D(E)}{v_{sh}^2},
\end{equation}
with $v_{sh}$ the velocity of the upstream fluid in the downstream frame.
By posing $t\approx T_{acc}(E_{max})$, one obtains:
\begin{equation}\label{eq:Emax_B}
E_{max}(t)\simeq \frac{E_{sh}}{3\kappa}\omega_c t,
\end{equation}
where we introduced $\kappa\equiv D(E_{max})/D_B(E_{max})$.

\begin{figure}\centering
\includegraphics[trim=30px 0px 40px 5px, clip=true, width=.485\textwidth]{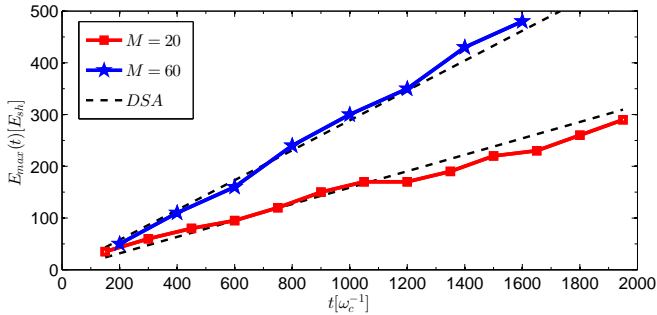}
\caption{\label{fig:Emax}
Time evolution of the maximum ion energy for parallel shocks with $M=20$ and 60 (Runs B and F in Table \ref{tab:box}), compared with the DSA prediction  according to Equation \ref{eq:Emax_B}, with $\kappa_{20}= 2.1$ and $\kappa_{60}=$1.2, respectively (dashed lines).}
\end{figure}

Figure \ref{fig:Emax} shows the time evolution of the inferred maximum ion energy for Runs B and F in Table \ref{tab:box}, corresponding to $M=20$ and 60. 
We compare $E_{max}(t)$ with the scaling provided by Equation~\ref{eq:Emax_B}, obtaining good fits with $\kappa_{20}\approx 2.1$ and $\kappa_{60}\approx 1.2$ for $M=20$ and $M=60$, respectively (dashed lines in Figure \ref{fig:Emax}).
For the low-$M$ case, $\kappa_{20}$ is consistent with the value of $D(E_{max})$ inferred from Figures \ref{fig:CoeffDiff20} and \ref{fig:cd2080}, within a factor of about 2.
As an example of a high-$M$ case, we follow for a long time a quasi-1D shock with $M=60$ (Run F), which is a reasonable choice because the particle spectrum is not very dependent on the transverse size of the simulations, as we showed in the appendix of Paper II.
Very interestingly, the value of $\kappa_{60}$ for the $M=60$ case is smaller than for the lower-$M$ shock, attesting to the relevance of magnetic field amplification (which increases with the shock strength, see Equation \ref{eq:db}) in favoring the rapid energization of accelerated particles. 
In particular, since 
\begin{equation}
\kappa=\frac{D(E_{max})}{D_B(E_{max})}\propto \frac{B_0}{B_{tot}}\propto \frac1{\sqrt{M}},
\end{equation}
one would expect $\kappa_{60}/\kappa_{20}\approx \sqrt{60/20}=\sqrt{3}$, which is in good agreement with the best-fitting values in Figure \ref{fig:Emax}.

It is worth remembering that Equation~\ref{eq:Emax_B} is expected to be accurate only within a factor of a few, since $\kappa$ is in principle a function of time, and both $D(E)$ and $u_1$ should be functions of the position in the precursor.
Yet, the good agreement between DSA theory and simulations confirms that the acceleration time is dominated by the most recent (longest) cycle, and suggests that diffusion is a good approximation for the transport of non-thermal ions up to the the exponential cut-off; 
this can also be viewed as an independent estimate of the diffusion coefficient close to $E_{max}$.
Moreover, we showed that the more effective the magnetic field amplification, the more rapid the increase of the CR maximum energy with time; 
this is a direct consequence of the scaling of the diffusion coefficient with $B_{tot}/B_0$, and has crucial implications on the maximum energy achievable in given classes of sources, and in particular on the possibility of producing PeV protons in SNRs.

As a final comment, we report that \cite{gs12} found a significantly shallower dependence of $E_{max}$ on $t$ after $t\sim 200 \omega_c^{-1}$ (figure 9 in their paper) because of the use of small computational boxes.
Instead, our large longitudinal dimension in run B ($10^5c/\omega_p$) allows us to properly account for the diffusion lengths of the most energetic ions until $t\approx 2000\omega_c^{-1}$.

\section{Conclusions}\label{sec:concl}
This paper is the third of a series aimed to investigate several aspects of ion acceleration at non-relativistic shocks through hybrid simulations. 
In previous papers \citep[][Paper I,II]{DSA,MFA}, we outlined the features of DSA acceleration and magnetic field amplification.
Here, we study the effects of self-generated magnetic turbulence on the accelerated particles, characterizing how particles diffuse in pitch-angle, which corresponds to a random walk in space.
Particle diffusion allows multiple shock crossing, and is crucial in regulating the acceleration rate. 
We find that, in the shock precursor of quasi-parallel shocks, accelerated ions are scattered by the self-generated magnetic turbulence, with a mean free path roughly comparable with the particle's gyroradius.
There are several interesting points to notice.
\begin{itemize}
\item At low Mach numbers ($M\lesssim 30$), scattering is due to resonant Alfv\'en waves with amplitude $\delta B/B_0\lesssim 1$ generated by accelerated ions (Figure \ref{fig:CoeffDiff20}), as predicted within the quasi-linear theory of streaming instability (section 3 of Paper II).
In this case, the \emph{self-generated diffusion} coefficient must be calculated in the fraction of the total magnetic field in waves with resonant wavelengths (Equation~\ref{eq:Dsg}).

\item For strong shocks, instead, $\delta B/B_0\gg 1$, and the wave spectrum has a more complicated shape because of the relevance of the NRH instability.
Most of the magnetic energy is in modes resonant in wavelength with high-energy ions (figure 7 in Paper II), and all the accelerated particles feel large-scale, non-linear perturbations. 
The result is that energetic particles experience \emph{Bohm-like diffusion} in the total (amplified) magnetic field (Figure \ref{fig:CoeffDiff80}).

\item We calculate the local diffusion coefficient: i) by exploiting the analytic theory of DSA (Section \ref{sec:anal}); and 
ii) by tracking individual particles in self-consistent electromagnetic fields (Section \ref{sec:track}). 
The two methods return very consistent results.
The latter allows us also to investigate the spatial dependence of the diffusion coefficient (Figure \ref{fig:cdx}).

\item The evolution of $E_{max}(t)$ is governed by the time it takes for energetic ions to diffuse back and forth across the shock. 
Bohm-like diffusion in the amplified field accounts reasonably well for such an evolution (Figure \ref{fig:Emax}), providing another independent test of particle diffusion at highest energies.

\item The maximum energy achievable in a given amount of time depends on magnetic field amplification (compare the curves for shocks with $M=20$ and 60 in Figure \ref{fig:Emax}): 
stronger shocks accelerate CRs up to larger energies, proportional to the suppression of the diffusion coefficient produced by the larger amplification of the initial magnetic field.

\end{itemize}

In a forthcoming publication we will cover the mechanisms that lead to the injection of ions into DSA, in order to provide closure for the present series of papers. 


\subsection*{}
We wish to thank L.\ Gargat\'e for providing a version of \emph{dHybrid}, and the referee for her/his precious comments.
This research was supported by NSF grant AST-0807381 and NASA grant NNX12AD01G, and facilitated by the Max-Planck/Princeton Center for Plasma Physics. 
This work was also partially supported by a grant from the Simons Foundation (grant \#267233 to AS), and by the NSF under Grant No.\ PHYS-1066293 and the hospitality of the Aspen Center for Physics.
Simulations were performed on the computational resources supported by the PICSciE-OIT TIGRESS High Performance Computing Center and Visualization Laboratory. This research also used the resources of the National Energy Research Scientific Computing Center, which is supported by the Office of Science of the U.S. Department of Energy under Contract No.\ DE-AC02-05CH11231, and XSEDE's Stampede under allocation No.\ TG-AST100035.

\bibliographystyle{yahapj}
\bibliography{diffusion}
\end{document}